\documentclass[twocolumn,showpacs,preprintnumbers]{revtex4}  
\usepackage{graphicx}
\usepackage{dcolumn}
\usepackage{bm}
\usepackage{ifthen}
\usepackage{booktabs}
\usepackage{amsmath}

\begin{document}

\title{Experimental demonstration of nonlocal effect in partial collapse measurement and reversal operation}
\author{Xiao-Ye Xu, Jin-Shi Xu, Chuan-Feng Li$\footnote{
email: cfli@ustc.edu.cn}$, Yang Zou, and Guang-Can Guo}
\affiliation{Key Laboratory of Quantum Information University of
Science and Technology of China, CAS, Hefei, 230026, China}

\begin{abstract}

We demonstrate experimentally the nonlocal reversal of a
partial-collapse quantum measurement of a two-photon entangled state.
Both the partial measurement and the reversal operation are
implemented in linear optics with two displaced Sagnac
interferometers, characterized by single qubit quantum
process tomography. The recovered state is measured by quantum state
tomography and its nonlocality is characterized by testing the Bell
inequality. Our result will be helpful in quantum communication and
quantum error correction.
\end{abstract}

\maketitle
Quantum measurement (QM) was firstly explained by von Neumann as a projective measurement \cite{von} and then developed to be a general measurement theory in standard quantum mechanics \cite{Dirac}. Different from other quantum process \cite{Nielsen00}, wave packet collapse in QM, which can monitor the transition from quantum to classical \cite{Zurek91}, is too rapid to test and depict \cite{Schlosshauer}. During the early days of QM, this collapse was just recognized as a postulate \cite{von, Dirac}. In the 80's, it was reinterpreted as a consequence of environment induced decoherence \cite{Zurek,Zurek91,Zurek03}. In the framework of decoherence theory, a complete QM process would be an evolution of an open system, containing three parts: a quantum system (S), a measurement apparatus (A), and an environment (E), that could be divided into two steps: an interaction between S and A that produces correlations between them, and a decoherence induced by E transforming a quantum measurement into a classical probable event \cite{Zurek03}. When considering all elements in a quantum measurement, the general quantum measurement theory will be reinterpreted in terms of positive operator valued measures (POVMs) \cite{Peres,Nielsen00}.

A special kind of POVM arises in the partial collapse measurement (PM) in which the initial state changes only weakly and yields little information about it \cite{Elitzur,Wootters79,Durr}. In this situation, although the interaction between S and A is not strong, the orthogonal decomposition of subsystem A and the decoherence of the composite system in E will partially collapse the measured quantum system. For the partial extraction of information from PM, a general QM process can be tested and characterized in detail by changing the interaction intensity in PM (defined in term of partial collapse strength $p$) \cite{Katz}. Unlike irreversible collapses in QM processes in general, a state collapse in a PM is unsharp and can be reversed, i.e., the system's initial state can be recovered \cite{Mabuchi,Nielsen,Raimond}. This state recovery is useful in quantum error corrections \cite{Koashi} and has been experimentally demonstrated in solid state systems \cite{Katz,Korotkov,Katz08} as well as in linear optics \cite{Kim}. Partial state collapses in PMs can also be utilized to simulate and analyze amplitude damping channels \cite{Nielsen00,Gisin}.

As a useful resource, quantum entanglement is very flimsy and subject to unavoidable degradation under environment induced dissipation and decoherence \cite{Mintert,Yu}. However, by exploiting quantum measurement, we can overcome quantum disentanglements in noisy channels \cite{Xu}. In addition, hidden entanglement can also be distilled by PM \cite{Kwiat01} and entanglement dynamics induced by PM in solid state systems has been studied \cite{Ge}. Elitzur and Dolev have pointed out \,\cite{Elitzur} that a PM has nonlocal effects; in more detail, PM decoherence on a single subsystem state of an entangled state can be compensated by a corresponding PM on the other subsystem state, by which means the initial entangled state can be recovered. Such nonlocal effects can be helpful in studying quantum paradox, such as Hardy's paradox \cite{Lundeen}, and in utilizing entanglement in real quantum communication channels. In this paper, we describe an experiment demonstrating these nonlocal effects and realizing a nonlocal reversal of PM in a two qubits photonic system.

\begin{figure}
\begin{center}
\includegraphics[width=3in]{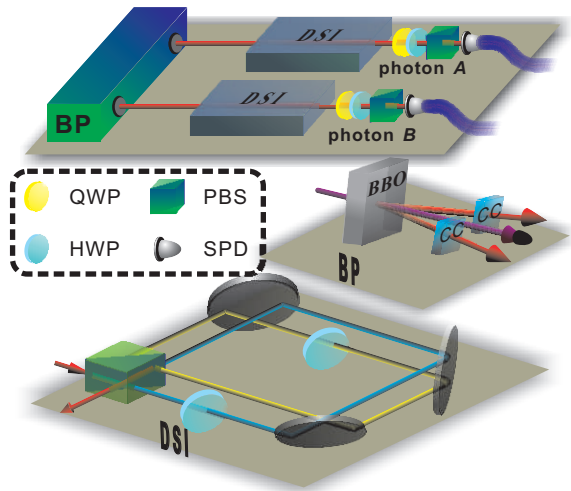}
\end{center}
\caption{(color online). Scheme of the experimental setup. Degenerate polarization-entangled photon pairs \cite{Kwiat99} of wavelength 800\,nm are generated by two 0.5-mm-thick $\beta$-barium-borate (BBO) crystals, cut at 29.8$^\circ$ and pumped by a mode locked Ti:sapphire laser. Using two compensatory crystals (CC), we erase the time difference and prepare as an initial state one of the Bell states. The partial measurement of the photon's polarization is realized by displaced Sagnac interferometer and two half wave plates (HWP) in each arm. The final state is reconstructed by quantum state tomography \cite{James} with two measurement devices composed of a quarter wave plate (QWP), a HWP, polarizing beam splitter (PBS), and a single photon detector (SPD). BP signifies Bell state preparation.}
\label{fig:setup}
\end{figure}

A essential part of a PM is the detector, that with probability $p$ collapses the system to a prescribed state (for example, a vertical polarization state denoted here by $|V\rangle$). Even if the detector is not click, failure of this probable detector can mean partial gain in information about the system and thus produces a physical collapse to a state that is non-orthogonal to one produced in the event of a successful case \cite{Elitzur}. The state evolution in a non-click event, usually defined as an interaction free measurement \cite{Elitzur93}, can be in theory represented by a measurement operator in the measurement basis of $|H\rangle$ (horizontal polarization) and $|V\rangle$
\begin{eqnarray}
P_M =
\begin{pmatrix}
1        & 0\\
0        &\sqrt{1-p}
\end{pmatrix}.
\end{eqnarray}
Unlike irreversible projective measurements, the matrix $P_M$ has a mathematical inverse
\begin{eqnarray}
R_M = \frac{1}{\sqrt{1-p}}P_M',
\end{eqnarray}
where $P_M' =
\begin{pmatrix}
\sqrt{1-p}       & 0\\
 0                &1
\end{pmatrix}$ corresponds to a partial measurement with the same partial collapse strength $p$ as $P_M$ on the orthogonal basis. $R_M$ is physically defined as reversal operator of partial measurement $P_M$ \cite{Koashi}. Here, we consider partial collapse measurements and reversal operations of the two photon Bell state $|\psi\rangle = \frac{1}{\sqrt{2}}(|HH\rangle+|VV\rangle)$ \cite{Bell}. A $P_M$ operation acting on the first photon (labeled by $A$ ), mathematically expressed as $P_M\otimes I$ ($I$ representing the identical operator here acting on the other photon $B$), will change $|\psi\rangle$ to
\begin{eqnarray}
|\psi_1\rangle =
\sqrt{\frac{1}{2-p}}(|HH\rangle+\sqrt{1-p}|VV\rangle),
\end{eqnarray}
the concurrence \cite{Wootters} of which is $2\sqrt{1-p}/(2-p)$. Subsequently, a local reversal of $P_M$ can be realized by a $R_M$ operation on photon $A$. The entire process can be mathematically expressed as $(R_M\cdot P_M)\otimes I$, which equals $I\otimes I$ if $R_M$ is of the same partial collapse strength as $P_M$ and the maximal entangled state $|\psi\rangle$ can be revived from $|\psi_1\rangle$, that is to say, partial information about initial state $|\psi\rangle$ obtained from the previous $P_M$ operation can be erased. However, as Elitzur and Dolev have pointed out \cite{Elitzur}, a nonlocal reversal of the previous $P_M$ can be realized by a $R_M$ operation on the second photon. Consequently, the entangled state $|\psi_1\rangle$ can be nonlocally revived to $|\psi\rangle$ by employing $R_M$ with the same collapse strength as $P_M$.

\begin{figure}
\includegraphics[width=3in]{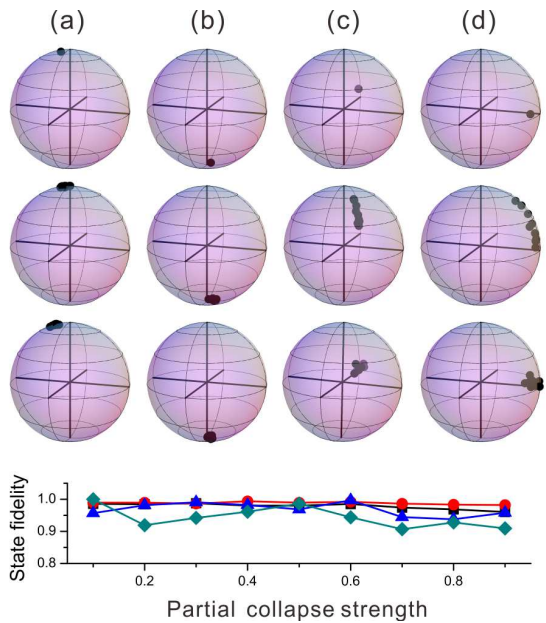}
\caption{(color online). Schematic diagram illustrating with the aid of the Bloch sphere the state evolution under partial collapse measurement and reversal. (a), (b), (c), and (d) represent the evolution for different input states ($|H\rangle, |V\rangle, |R\rangle$, and $|D\rangle$ respectively). The first row represents the initial states, the second and third row represent the state after partial measurement and reversal. The last row depicts the fidelities of recovered states after reversal. $|H\rangle, |V\rangle, |R\rangle$ and $|D\rangle$ states are represented by square, circle, triangle and diamond. Error bars are smaller than the spot size.}
\label{fig:sigle state}
\end{figure}

The $P_M$ and $R_M$ operations can be implemented with a set of Brewster-angle glass plates \cite{Kwiat01,Kim}, the collapse strength of which can only be adjusted within a limited range (0.4-0.9) and certain discrete values. In our experiment (see Figure\,\ref{fig:setup} for setup details), these are implemented with two Sagnac interferometers \cite{Almeida} and four HWPs, which can manipulate independently and coherently the photon's polarization with high fidelity. Each interferometer has two HWPs, one in $|V\rangle$ path (labeled by HWP1) and the other in $|H\rangle$ path (labeled by HWP2). The partial collapse strength $p$ is given by $\sin^2\theta$, where $\theta$ is the angle of the HWPs, and in our experiment can be adjusted continuously within the full range (0, 1). The operation $P_M$ with collapse strength $p$ is realized by rotating HWP1 through angle $\theta$ in photon $A$'s interferometer. A local reversal is realized by rotating HWP2 through the same angle as HWP1 in photon $A$'s interferometer. A nonlocal reversal is realized by rotating HWP2 through the same angle as HWP1 in photon $B$'s interferometer.

\begin{figure}
\begin{center}
\includegraphics[width=3in]{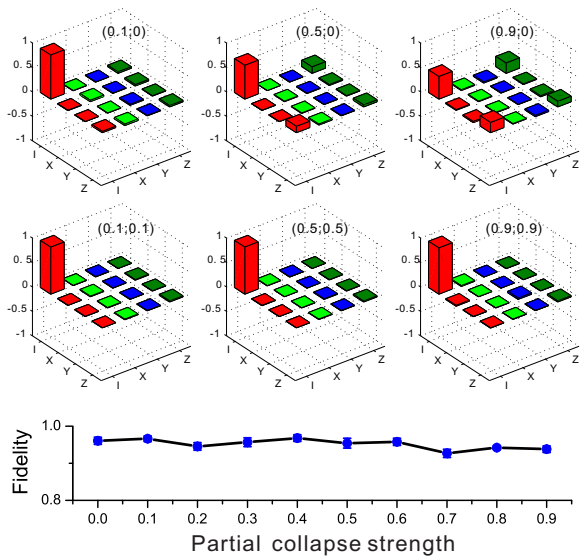}
\end{center}
\caption{(color online). Real parts of $\chi$ matrix in the quantum process tomography for: First row, partial collapse measurement; Second row, reversal of previous partial measurement. (i,j) in each graphic represents the partial collapse strength of $P_M$ and $R_M$ respectively. The imaginary parts of the $\chi$ matrices are negligible. The third row shows the fidelity of both the partial collapse measurement and reversal operation together at different partial collapse strengths.} \label{fig:process}
\end{figure}

Generally, quantum processes can be described as a quantum state evolution (QSE) of a system, that is, tomography of the state after evolution for a designated input state. In Fig.\,\ref{fig:sigle state}, we have schematically illustrated the state evolution of four input states ($|H\rangle,|V\rangle,|R\rangle=(|H\rangle-i|V\rangle)/\sqrt2,$ and $|D\rangle=(|H\rangle+|V\rangle)/\sqrt2$) of a heralded single photon with the aid of the Bloch sphere. With increasing partial collapse strength, states $|R\rangle$ and $|D\rangle$ will gradually collapse to state $|H\rangle$. As long as the collapse is not completely to $|H\rangle$, the initial states can be recovered. In the last row of Fig.\,\ref{fig:sigle state}, we depict the fidelity of the recovered state \cite{Barnum} obtained by the reversal operation. The slightly depressed fidelities for cases $|R\rangle$ and $|D\rangle$ are caused by the phase error due to the HWPs in the Sagnac interferometer, which can be corrected by a phase compensate plate \cite{Bongioanni}.

\begin{figure}
\includegraphics[width=3in]{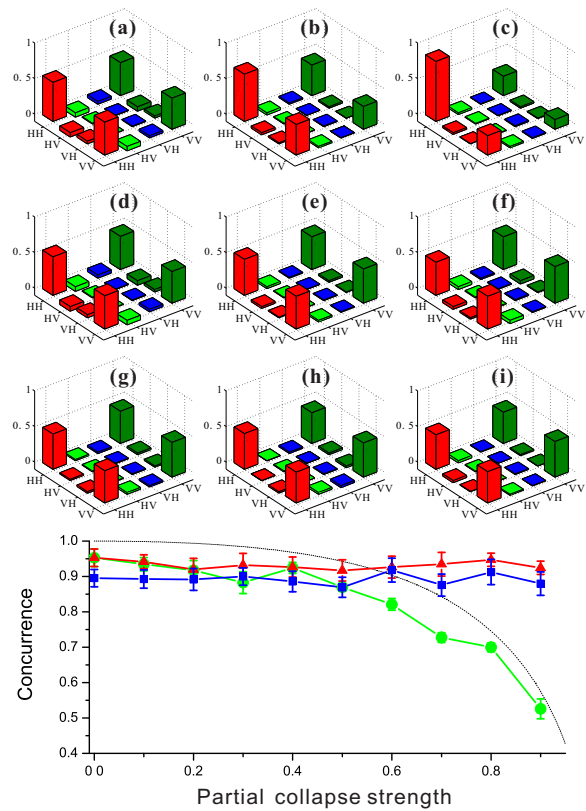}
\caption{(color online). Schematically diagrams illustrating the entangled state evolution under local and nonlocal reversals of the partial collapse measurements using the real parts of density matrices. The imaginary parts are negligible. (a), (b), and (c) depict the states with partial collapse strengths 0.1, 0.5, and 0.9 respectively after partial measurements. (d), (e) and (f) depict the recovered states of those in (a), (b), and (c) by local reversal with the same partial collapse strength accordingly. (g), (h), and (i) depict the recovered states of those in (a), (b), and (c) by nonlocal reversals. The last row graphs the concurrences of states after partial measurement (circle), local reversal (triangle) and nonlocal reversal (square). The dashed line represents the theoretical prediction of the concurrence of the state after partial measurement calculated from Eq.\,(4).} \label{fig:entangled state}
\end{figure}

Next, we use quantum process tomography (QPT) \cite{Nielsen} to fully characterize this single qubit operator $P_M$ and  $R_M\cdot P_M$, implemented by using Sagnac interferometers. In the QPT method, $P_M$ and $R_M\cdot P_M$ can be fully represented by a $\chi$ matrix in the basis $(I,X,Y,Z)$, where $X,Y,Z$ represent the Pauli matrices. From Eq.\,(1), the $\chi$ representation of $P_M$ is $[(1+\sqrt{1-p})I+(1-\sqrt{1-p})Z]/2$, reducing to $I$ when $p=0$ and $(I+Z)/2$ in the limit $p\rightarrow1$. For a reversal operation with the same partial collapse strength intended to recover the initial state from the state after partial measurement, the
effect of process $P_M$ plus $R_M$ equals an identical operation $I$. In the experiment, we reconstructed the $\chi$ matrices of $P_M$ and $R_M\cdot P_M$ by using the maximum likelihood estimation method \cite{Fiurasek}, which are shown in the first and second rows of Fig.\,\ref{fig:process}. Taking $P_M$, it is clear that, with
increasing partial collapse strengths $p$, the coefficient of $I$ decreases and the effects of the $Z$ operation begin to emerge. For reversals, these equal $I$ for different partial collapse strength within the range of error. To quantitatively analyze the reversal of the partial measurement on a single qubit, we calculate the reversal fidelity \cite{Barnum} defined as
\begin{eqnarray}
F_{process} = Tr[\sqrt{\chi_I^{1/2}\chi_{exp}\chi_I^{1/2}}]^2,
\end{eqnarray}
where $\chi_I$ represents the $\chi$ matrix of $I$ and $\chi_{exp}$ represents the $\chi$ matrix reconstructed from the experiment. In the third row of Fig.\,\ref{fig:process}, we depict the reversal fidelity for different partial collapse strengths. Generally, the $P_M$ and $R_M$ operations of a single qubit should be implemented by two independent elements. In our experiment, these are implemented within one interferometer. For this reason, the reversal fidelity is above $0.93 \pm 0.01$, indicating that the visibility of our Sagnac interferometers is large enough to implement a nonlocal reversal operation.

After characterizing the $P_M$ and $R_M\cdot P_M$ operations of single qubit, we study the local and nonlocal reversal of partial collapse measurement on the entangled state by using QSE analysis. A partial measurement on state $|\psi\rangle$ changes it to $|\psi_1\rangle$, these are graphically represented in (a), (b) and (c) of Fig.\,\ref{fig:entangled state}. The local reversal of partial collapse measurements with the same partial collapse strength can revert the system to its initial state, graphically represented in (d), (e) and (f) of Fig.\,\ref{fig:entangled state}. The nonlocal reversal of partial measurements with the same partial collapse strength has the same restorative effects and these are graphically represented in (g), (h) and (i) of Fig.\,\ref{fig:entangled state}. In the last row of Fig.\,\ref{fig:entangled state}, the concurrence of the states after partial measurement, local reversal, and nonlocal reversal are depicted. Because the initial state prepared in the experiment is not pure enough (the measured concurrence is $0.95 \pm 0.02$), the concurrences of the states after partial measurement are all below
the theoretical prediction (dot dashed line in last row of Fig.\,\ref{fig:entangled state}). However, these have similar variational tendencies predicted by theory. Since only one interferometer is used to implement local reversals, the states after partial measurement can be recovered with a large concurrence, ($0.92 \pm 0.03$ for the worst case) by using local reversal. In nonlocal reversals, the states can be recovered with concurrence around 0.9. Large concurrences of the recovered states predict a high interference visibility, that is, 0.96 for a single interferometer and 0.91 for two interferometers.

Another important characteristic of quantum mechanics is nonlocality \cite{Bell}, which will be a very useful feature to exploit in quantum technologies \cite{Gisin09}. In our experiment, we measured the nonlocality of recovered states by the Clauser-Horne-Shimony-Holt inequality, $S\leq 2$ for any local realistic theory \cite{Clauser}. By calculating the maximal value of $S$ from the measured density matrix for a recovered state at partial collapse strength $p = 0.5$, we get the maximal violation angle $(\theta_1 = -3.6^\circ, \theta_1' = -18.0^\circ, \theta_2 = -46.8^\circ, \theta_2' = 21.6^\circ)$. Accordingly, the measured value of S is $2.538 \pm 0.035$, which violates the local realism limit 2 by over 15 standard deviations.

In conclusion, by using inherently stable modified Sagnac interferometers with high interference visibility, we have experimentally demonstrated local and nonlocal reversals of partial measurements on photonic entangled state.
Our result will be helpful in quantum communication and quantum error correction.


This work was supported by National Fundamental Research Program and
the National Natural Science Foundation of China (Grant Nos.60921091,
10874162, 10734060 and 11004185).

\end{document}